\documentstyle[aps,preprint]{revtex}
\tightenlines
\begin{document}

\title{Dynamics  of a  driven probe molecule
in a liquid monolayer}

\author{J.De Coninck$^{1}$, G.Oshanin$^{1,2}$ and  
M.Moreau$^{2}$}

\address{
     $^{1}$ Centre de Recherche en Mod\'elisation Mol\'eculaire, 
Service de Physique Statistique
 et Probabilit\'es, Universit\'e de
Mons-Hainaut, 20 Place du Parc, 7000 Mons, Belgium\\
     $^{2}$ Laboratoire de Physique Th\'eorique des Liquides, 
Universit\'e Paris VI, 4 Place Jussieu, 75252 Paris Cedex 05, 
France}

\vspace{0.2cm}

\address{\mbox{ }}
\address{\parbox{14cm}{\rm \mbox{ }\mbox{ } 
We study dynamics of a probe molecule, driven by an
 external constant force
in a liquid monolayer on top of solid surface. In 
terms of a microscopic, 
mean-field-type approach,   
we calculate the terminal
velocity of the probe molecule. This allows us to
establish the analog of the Stokes formula, 
in which the friction
 coefficient is interpreted
in terms of
the microscopic parameters
characterizing the system. We also determine the
distribution of the monolayer
particles as seen from the stationary moving 
probe molecule
and
estimate the self-diffusion coefficient
for diffusion in a liquid monolayer.
}}
\address{\mbox{ }}

\vspace{4cm}
\address{\parbox{14cm}{\rm PACS No: 68.15+e; 
05.60+w}}
\maketitle

\makeatletter
\global\@specialpagefalse

\makeatother

\vspace{1cm}

Thin liquid films adsorbed  on solid surfaces or
confined in a narrow space between two solids represent
a remarkable example of a two-phase physical system,
 in which an
intrinsically disordered 
liquid phase is spanned by
and contends with the ordering potential of the
solid.
It is well appreciated by now
 that the behaviour of such  films 
is markedly 
different of the customary 
behaviour of the bulk liquids:
experiments reveal 
effects of solid-like or glassy response 
 to an
external shear, sharp
 increase of 
relaxation times and of the films' viscosity  
\cite{isra,gran,ciep,krim2,joel}.  The 
departure from the bulk behaviour
is progressively more
pronounced the thinner the film is and 
is interpreted usually as
the manifestation of intricate
 cooperative effects, whose nature is not 
yet completely elucidated (see, e.g. 
\cite{gran}). 

\vspace{0.2cm}

In this Letter we study cooperative behaviour 
in a liquid monolayer on solid surface,
 arising in response 
to internal perturbances created by a single
probe molecule (say, a charge carrier), which is driven
by constant external force.  
We propose here a simple analytically 
soluble  model,
 which  allows us to determine
the stationary, non-homogeneous distribution of
 the monolayer particles as seen from the 
probe molecule,
 and   
to derive a closed-form
 non-linear equation relating
 the terminal velocity of the probe
to the magnitude of the driving force $F$.
We find that in the limit of
small $F$ this equation
reduces to the  
Stokes viscous-flow law, in which
 the 
 friction coefficient $\xi$ is determined
explicitly, in terms of the microscopic
 parameters
characterizing the system under study.
 We show that $\xi$ combines 
 the contributions
of two different 
"dissipative" processes 
 - temporary trapping
of the probe
by the potential wells, due to the interactions
with the solid atoms, and the cooperative 
effects associated
with the formation of a non-homogeneous 
distribution 
of the monolayer particles around the probe.
The knowledge of $\xi$ also 
allows us  to 
estimate 
the self-diffusion coefficient for 
diffusion in a liquid monolayer.

\vspace{0.2cm}

We start with the formulation of our model.
Consider a 2D homogeneous 
monolayer of identical interacting
particles moving randomly on top of an 
ideal crystalline surface (Fig.1). 
We suppose first
that the particles' motion 
is induced by chaotic vibrations 
of the solid atoms. 
Second, we assume
that the particle-particle interactions 
(PPI) are essentially 
weaker than the particle-solid interactions 
(PSI), such that
the former don't perturb the wafer-like
 array of potential wells
created due to the PSI (Fig.1). 
This permits us to adopt the conventional picture 
of the dynamics under such conditions
 (see \cite{adam} and references therein):
 the particles migration proceeds
by thermally activated rare events of hopping from
one well to another in its neighborhood. 
The hopping events are separated
by the time interval $\tau^{*}$, which is 
the time a
 given particle typically
spends in each well vibrating around its minimum; $\tau^{*}$ 
is related to the temperature $\beta^{-1}$ 
and the barrier of the
PSI by the Arrhenius formula. 
Further on, the PPI couple the dynamics of a
given particle to motions of
 all others. That is, 
a particle escaping from a given well 
follows preferentially the gradient of the
PPI
potential (see, e.g. \cite{leba,tur} for discussion). 
The realistic  PPI are repulsive at
short scales and show 
a weak, long-range attraction at
longer interparticle separations.  
Here we simplify the actual behaviour by
neglecting, first, the
particle-particle attractions and, second, by
approximating the
repulsive part of the interaction
potential by an abrupt hard
wall, imposed at the distance equal to
 the interwell spacing. 
In such an idealized model
the hopping
probabilities are naturally
 decoupled of the particles
 distribution
 and all hopping directions
 are equaly probable. 
The hard-core repulsion 
 prevents multiple occupancy
of any potential well - we suppose
 that a particle may hop 
into the target well only in case
 when the latter is 
unoccupied at this moment of time; otherwise, the particle
attempting to hop is repelled back to its position.

\vspace{0.2cm}

The key aspect in our model is the 
 probe molecule (abbreviated as the PM in the following), 
which is subject to an external constant (small)
force $F$ directed
along the $X$-axis (Fig.1).
We suppose that
the PM also
performs an activated hopping motion, 
constrained by the hard-core
 interactions with the monolayer particles. 
The typical time which the PM spends
being trapped by each well is denoted as $\tau$, 
which is dependent on the 
probe-solid 
interactions and hence may differ from $\tau^{*}$.  
Varying  $\tau$
we can mimic different possible regimes. 
For instance, the choice $\tau = 0$
corresponds to the situation when  
the PM simply slides regardless 
of the surface corrugation.  
Now, in contrast to the monolayer particles the  
 PM dynamics is
anisotropic due to the
 applied force; that is, on escaping
 from the wells,
the PM attempts to hop preferentially in the direction of $F$.
 Supposing for simplicity 
 that the wells form a
 square $(X,Y)$-lattice of spacing $\sigma$, we define 
 the probability $q(X,Y|X',Y')$ 
 of an attempt to hop 
from the well at  $(X,Y)$ to 
the adjacent well at  $(X',Y')$ as:
\begin{equation} \label{eqn1} 
q(X,Y|X',Y')  \; = \; Z^{-1} \; 
\exp( \beta \;  F \; (X' \; - \; X)/2), \; Z = 4 \; \cosh^{2}(\beta \sigma F/4) 
\end{equation}
To take into account the
hard-core repulsion between the PM and the monolayer particles 
we stipulate that the PM hops into 
the target well 
only in the case when it is vacant at this moment of time.
Otherwise, it remains at its position.

\vspace{0.2cm}

Let $P(X,Y)$ ($\rho(X,Y)$)  
denote  the probability that at time $t$ the PM 
(a monolayer particle)
occupies
the well with the
 coordinates $(X,Y)$. 
To derive the closed-form equations
 describing the time evolution of
these properties, we assume that 
 correlations between the
occupations of different wells decouple.
It was shown recently in \cite{burb} that
 such a decoupling  
provides an adequate description 
of the  driven particle  dynamics in a 
one-dimensional symmetric 
hard-core lattice gas, in which case the
 hindering effect
of the gas particles is much more strong 
than in 2D. We thus expect
that in the 2D situation under study such
 an approach will yield
correct dependence of the PM terminal 
velocity on the system
parameters; we do not expect, of course,
 to obtain exact 
 numerical factors.
Decoupling the correlations, 
 we have 
\begin{equation} \label{eqn2} 
4 \; \tau^{*} \; \frac{d \rho(X,Y)}{d t}  \; = 
\; \sum_{(X',Y')} \{ - \; 
(1 \; - \; \rho(X',Y')) \; 
 \rho(X,Y) \; 
+  \; (1 \; - \; \rho(X,Y)) \;  \rho(X',Y')\},
\end{equation}
which holds  for all wells excluding 
the wells adjacent to the PM position.
Next, we find 
\[\tau \; \frac{d  P(X,Y)}{d t}  \; = \; \sum_{(X',Y')} 
\{ - \; q(X,Y|X',Y') \;  P(X,Y)  \; 
(1 \; - \; \rho(X',Y')) \; + \]
\begin{equation} \label{eqn3}  
 + \;  q(X',Y'|X,Y) \; 
 (1 \; - \; \rho(X,Y)) 
 \; P(X',Y')\}
\end{equation}
The sums in eqs.(\ref{eqn2}) and (\ref{eqn3}) run 
over all wells $(X',Y')$ adjacent to 
the well at position 
$(X,Y)$.

\vspace{0.2cm}

Multiplying both sides of eq.(\ref{eqn3}) by $X$
and summing over all wells we obtain
\[V \; =  
\; \frac{\sigma}{\tau} \; (q_{+} \; - \; q_{-}) \; - 
 \; \frac{\sigma}{\tau} \;
\sum_{X,Y} P(X,Y) \; 
 (q_{+} \; \rho(X + \sigma, Y) \; - 
  \; q_{-} \; \rho(X - \sigma, Y)) \; = \]
\begin{equation} \label{eqn4}
= \;  \frac{\sigma}{\tau} \; [q_{+} \; (1 \; - \; 
 f_{V}(\lambda_{1} = \sigma, \lambda_{2} = 0)) \; 
- \; 
 q_{-} \; (1 \; - \;  f_{V}(\lambda_{1} = - \sigma, 
\lambda_{2} = 0))], 
\end{equation}
where $V$ is the $X$-component of the PM velocity, 
$q_{\pm} = \exp(\pm \beta \sigma F/2)/Z$
and
 $f_{V}(\lambda_{1}, \lambda_{2})$ 
denotes the time-dependent probe-particle 
correlation function; the subscript $"V"$
 signifies that this property
is itself  dependent on the PM velocity. 

\vspace{0.2cm}

Dynamics
 of $f_{V}(\lambda_{1}, \lambda_{2})$
 is a superposition of two different processes - 
the particles diffusion  and the motion of the PM. 
From eqs.(\ref{eqn2}) and (\ref{eqn3}) we find that  
in the  continuous-space limit   
$f_{V}(\lambda_{1}, \lambda_{2})$ obeys
(see \cite{burb} for details of derivation in the 1D case):
\begin{equation} \label{eqn5}
\frac{\partial }{\partial t} \; f_{V}(\lambda_{1}, \lambda_{2}) \; = 
\; D^{*} \; (\frac{\partial^{2}}{\partial \lambda_{1}^{2}} \; + \; 
\frac{\partial^{2}}{\partial \lambda_{2}^{2}}) \; 
 f_{V}(\lambda_{1}, \lambda_{2}) \; + 
  \; V 
\; \frac{\partial }{\partial \lambda_{1}} 
\; f_{V}(\lambda_{1}, \lambda_{2}),
\end{equation}
where $D^{*} = \sigma^{2}/4 \tau^{*}$ is the "bare"
 diffusion coefficient of an isolated
 particle.

\vspace{0.2cm}

Let us discuss now the initial and boundary 
conditions to eq.(\ref{eqn5}).   
First, we suppose 
\begin{equation} \label{eqn6}
\left . f_{V}(\lambda_{1}, \lambda_{2})\right |_{\lambda_{1}, \lambda_{2}  \to \pm \infty} 
\; = \;
\left . f_{V}(\lambda_{1}, \lambda_{2})\right |_{ t = 0} \; = \; \rho, 
\end{equation}
which mean that  correlations between $P(X,Y;t)$ and 
$\rho(X + \lambda_{1}, Y + \lambda_{2})$ vanish in the limit
$\lambda_{1}, \lambda_{2} \to \pm \infty$ and that 
initially the  particles
 were uniformly distributed in 
the potential wells with mean density $\rho$.
Another set of four equations can be derived by 
analyzing the 
 evolution of $f_{V}(\lambda_{1}, \lambda_{2})$ in the 
 wells adjacent to
 the PM position (see for more details \cite{burb}). 
This gives the zero-flux boundary condition 
for $\lambda_{1} = 0, \lambda_{2} = \pm \sigma$ and
\begin{equation} \label{eqn7}
\left .  \;  \frac{\partial }{\partial \lambda_{1}}
 f_{V}(\lambda_{1}, 0)\right |_{\lambda_{1} = \pm \sigma} \; = \; - \; \frac{V}{D^{*}} 
\; f_{V}(\pm \sigma, 0) 
\end{equation}

Solution of the mixed boundary value
 problem in eqs.(\ref{eqn5}) to (\ref{eqn7}) yields the following result
for the particle distribution around the stationary moving PM:
\[f_{V}(\lambda_{1}, \lambda_{2}) \; = \;  
 \rho \; + \; 2  \;  \rho \; \exp( - \frac{\it{P} \lambda_{1}}{\sigma}) \; 
 \{
\frac{\rm{cosh}(\it{P})  -  \rm{2} \;  S_{-}  }{S_{+}  +  2  
S_{-}} \;  \sum_{n = - \infty}^{\infty}    \frac{I'_{\rm{2 n}}(\it{P})}{K'_{\rm{2 n}}(\it{P})} \; 
 T_{\rm{2n}}(\frac{\lambda_{1}}{r}) \;
 K_{\rm{2 n}}(  \frac{\it{P} {\rm  r}}{\sigma}) \; - \] 
\begin{equation} \label{eqn8}
 -  \;  \rm{2} \;   (1  +  \frac{ cosh(\it{P})  -  \rm{2}   S_{-}  }{\rm{4} \; 
(S_{+}  +  2   S_{-})}) \; 
  \sum_{n = 1}^{\infty} 
\; \frac{I_{2n - 1}(\it{P})}{K_{\rm 2n - 1}(\it{P})}
T_{2n - 1}(\frac{\lambda_{1}}{{\rm r}}) \; K_{{\rm 2 n - 1}}( \frac{\it{P} {\rm r}}{\sigma})\}, 
\end{equation}
where $S_{\pm}$ are given by
\begin{equation}  \label{eqn9}
S_{+}  \; = \;   -
 \;  \sum_{n = - 
\infty}^{\infty} \frac{I_{\rm{2 n}}'(\it{P})}{\{ ln[K_{\rm{2 n}}(\it{P})]\}' } \; 
\text{and} \;  
S_{-} \; = \; \sum_{n = 1}^{\infty} I_{\rm{2 n - 1}}(\it{P}) \; \{ ln[K_{\rm{2 n - 1}}(\it{P})]\}' 
\end{equation}
In eqs.(\ref{eqn8}) and (\ref{eqn9}) 
$\it{P}$ is the "Peclet number",  $\it{P} = V_{\infty} \sigma/{\rm 2 D^{*}}$,
$V_{\infty}$ is the PM terminal velocity, 
the prime stands for the derivative with respect to $\it{P}$, 
${\rm r} = \sqrt{\lambda_{1}^{2} + \lambda_{2}^{2}}$, while 
 $K_{n}(x)$, $I_{n}(x)$ and $T_{n}(z)$ 
 denote the modified Bessel functions and the 
Tchebyscheff polynomials respectively.   
Inserting eqs.(\ref{eqn8}) and (\ref{eqn9}) 
 into  eq.(\ref{eqn4}) we arrive at a closed with respect to $V_{\infty}$
 equation. This  non-linear transcendental equation  is not
analytically soluble,  and we will thus seek 
an approximate solution,  assuming 
that $V_{\infty}$ is sufficiently small, 
such that $\it{P} < {\rm 1}$. Expanding  $f_{V}(\pm \sigma,0)$
 to 
the second order in powers of
$\it{P}$ we  obtain 
\[V_{\infty}  \; \approx \; \frac{( q_{+} \; - \; q_{-}) 
 (1 \; - \; \rho)}{(\tau/\sigma) 
\; + ( q_{+} \; + \; q_{-}) \sigma
\rho/D^{*}} \; \]
\begin{equation} \label{eqn10}
 \times \;   [ 1 \;  
 + \; \frac{ 2 \rho}{1 - \rho} \; (\frac{( q_{+} \; - \; q_{-}) 
 (1 \; - \; \rho) \sigma^2}{ 2 D^{*} \tau
+ 2 ( q_{+} \; + \; q_{-}) \rho})^{2} \; 
 \{ \ln(\frac{( q_{+} \; - \; q_{-})  (1 \; - \; \rho) \sigma^2}{ 2 D^{*} \tau
+ 2 ( q_{+} \; + \; q_{-}) \rho}) \; +
\; \gamma \; + \; 1/2\}], 
\end{equation}
where 
 $\gamma = 0.577...$ is the Euler constant. 

\vspace{0.2cm}
 
Now, several comments on the result in eq.(\ref{eqn10})  are in order.  
We find, first, that $\it{P} < \rm{1}$ and 
our approximation is justified when the 
inequality $(q_{+} - q_{-}) 
(1 - \rho) \sigma^{\rm 2} < {\rm 2} D^{*} \tau \; + 
\; {\rm 2} (q_{+} + q_{-})
\sigma^{\rm 2} \rho$
is fulfilled.  This happens, namely, in the case 
when the applied force is sufficiently small, 
or when either $\rho > 1/3$, or $D^{*} \tau/\sigma^{2} > 1$. 
We note also that in the limit  $\it{P} < {\rm 1}$ the  second term 
in brackets in eq.(\ref{eqn10}) is much smaller than
unity and can be safely neglected. 
Further on, eq.(\ref{eqn10}) 
shows  that the PM velocity is controlled by
the combination of 
two different  factors - trapping by the wells 
and the hindering effect of the monolayer
particles.  
The first factor dominates in the limit
 when $D^{*} \tau/\sigma^{2} \gg \rho$, which
behaviour may take place in the case
when the barrier of the probe-solid interactions is 
much stronger than that for the
PSI.  
In this limit
eq.(\ref{eqn10}) reduces to  the essentially mean-field result 
$V_{\infty} \approx (q_{+} - q_{-}) \sigma/\tau_{\omega}$, 
where $\tau_{\omega} = \tau/(1 - \rho)$. 
We remark that since $\tau$ 
is appropriate for the motion in 
the stagnant layer,  rather 
than in the bulk phase, we may expect that even in this
 limit of 
the well-stirred monolayer the terminal 
velocity $V_{\infty}$
 will be
much less than the
corresponding 
(to this value of $F$) velocity in 
the bulk liquid. 
Next, the prevalence of the cooperative effects
appears in the limit when the ratio  
$D^{*} \tau/\sigma^2$ is less than $\rho$. 
Here we
 have 
$V_{\infty} \approx (q_{+} - q_{-}) (1 - \rho) D^{*}/ \sigma
\rho$, i.e. the terminal velocity is 
 proportional
to the "bare" diffusion constant of the monolayer
 particles and vanishes when $D^{*} \to 0$. 
This means that the  particles have to
 diffuse away of the PM in order it can move.
As a matter of fact, 
such a 
dependence mirrors an even stronger effect: 
the monolayer particles,
whose motion is random with no
preferential direction,  
 accumulate in front of the
 obstacle (the PM), which moves at a constant velocity. 

\vspace{0.2cm}

Next, 
using the notations of eq.(\ref{eqn1}) 
and  supposing that $\beta \sigma F \ll 1$ we cast
eq.(10) into the following physically revealing form
\begin{equation}  \label{eqn11}
F \; = \xi \; V_{\infty},  \; 
\xi \; = \; \frac{2}{(1 - \rho) \sigma \beta} 
\; (\frac{2 \tau}{\sigma} \; + \; \frac{\sigma
\rho}{D^{*}}),   
\end{equation}
which can be thought of as the analog of the Stokes
 formula for the system under study;
the friction coefficient $\xi$ is expressed through
 the microscopic parameters 
and displays the combined hindering effect of the
 potential
 wells and of the monolayer particles on
the PM dynamics. The friction coefficient is 
proportional to $\tau$, which
is related to the probe-solid interactions, and diverges
 when either $D^{*} \to 0$  or $\rho \to 1$. 

\vspace{0.2cm}

Further on, from the Einstein relation 
between the mobility and the diffusivity
of a test particle (the PM) 
we find
\begin{equation} \label{eqn12}
D_{PM} \; = \; \frac{(1 - \rho) \; D \; D^{*}}{D^{*} \; + \; 2 \; \rho \; D}
\end{equation}
where $D = \sigma^2/4 \tau$ and $D_{PM}$ are the PM
 "bare" and self-diffusion  coefficients
respectively.

\vspace{0.2cm}

Let us now briefly discuss 
the density distribution of the monolayer
particles as seen from  the stationary 
moving PM.  The local density in the nearest to the PM wells 
$f_{V}(\pm \sigma,0) 
 \approx   \rho  ( 1 \; \pm 
V_{\infty} \sigma/D^{*} )$, 
i.e. in front of (past) the PM the local density
 exceeds (is less than) 
the average 
density of the monolayer particles.
Further on, 
at large separations 
of the PM, such that $\lambda_{1} \; \gg \; 
 \Lambda =  D^{*} / V_{\infty}$, the density
profile is characterized by
\begin{equation} \label{eqn13} 
f_{V}(\lambda_{1}, 0) \; \approx  \; 
\rho \; ( 1 \; + \; (\frac{\sigma}{\Lambda})^{2} \; 
(\frac{\pi \Lambda}{\lambda_{1}})^{1/2} \; exp( -
\lambda_{1}/\Lambda)), 
\end{equation}
which means that  the excess density in front of the PM  
vanishes exponentially with the distance. 
The parameter $\Lambda$ can be identified as the 
characteristic length 
of the condensed region in front of the PM.  
We note, however, that this parameter 
is not very informative; particularly,
it diverges when either $D^{*} \to  \infty$ or $\rho \to 1$,
 which is a bit misleading
since in this limit the excess density tends to zero, 
i.e. the monolayer is homogeneous. 
A better parameter is
the integral excess density $\Omega$ along the $\lambda_{1}$-axis,
\begin{equation} \label{eqn14} 
 \Omega \; = \;  \int^{\infty}_{\sigma} d\lambda_{1} 
\; (f_{V}(\lambda_{1},0) \; - \; \rho)
\; \approx \; 
 \frac{(q_{+} - q_{-})  \rho ( 1 - \rho) \sigma^{3}}{D^{*} \tau  
+ (q_{+} + q_{-}) \sigma^{2} \rho} \;   
\{ - ln(\it{P}/\rm{2}) + 1 
- \gamma\} 
\end{equation}        
Eq.(\ref{eqn14})  shows that the condensed region is actually absent  
when either $\rho \to 1$ or $D^{*} \to \infty$ and increases when
  the driving force is increased. We also 
remark  that when the parameter $D^{*} \tau/\sigma^{2}$ 
is sufficiently large,
$\Omega$ is a non-monotoneous function of $\rho$. 

\vspace{0.2cm}

Finally,  we find that at large distances  past the 
PM the density follows
\begin{equation} \label{eqn15}
f_{V}(\lambda_{1}, 0) \; \approx \; \rho \; (1 \; + \; \frac{1}{8} \; 
(\frac{\sigma}{\Lambda})^{4} \; 
(\frac{\pi \Lambda}{|\lambda_{1}|})^{1/2}), \; |\lambda_{1}| \gg  \Lambda  
\end{equation}
Eq.(\ref{eqn15}) displays two remarkable features: First, the
particle density past the PM approaches its average
 value only as a power-law, 
which mirrors strong memory effects of the medium 
and signifies  that diffusive homogenization
of the monolayer, constrained by the hard-core interactions, 
 is a very slow process. Second, it shows that 
the density  is a non-monotoneous
 function of $\lambda_{1}$; $f_{V}(\lambda_{1} = - \sigma, 0) < \rho$ 
 and
 approaches $\rho$ from above when $\lambda_{1} \to - \infty$.

\vspace{0.2cm}

To summarize, we have studied dynamics of a probe molecule
driven by an external constant force in
a 2D monolayer of particles moving randomly
 on top of solid surface. We have proposed a microscopic
model description of such a system evolution and calculated 
 the PM terminal velocity $V_{\infty}$, self-diffusion coefficient $D_{PM}$ and 
the stationary 
distribution of the monolayer particles as seen from the probe molecule.  
We have shown that this distribution is strongly asymmetric: 
Past the PM, 
the local density is lower than the average density $\rho$ 
and tends to $\rho$ as a power-law of the distance, 
revealing strong memory effects of the
medium. In front of the PM the density is higher than the average -
the monolayer particles
accumulate 
in front of the  PM, creating a sort of a "traffic
 jam" which impedes its
motion.  We show that the PM terminal velocity
is controlled by the size of the jammed
region, which is, in turn,  dependent on
 the driving force,
 as well as
on the density of the monolayer
and on the rate at which 
 the monolayer particles may diffuse away of the PM. 
The balance between the driving force and the rate of
 the monolayer homogenization  
 manifests itself as a 
medium-induced frictional force exerted
 on the PM, which shows for small driving forces a
viscous-like behaviour. 
This resembles in a way the situation described in \cite{elie}, which
work considered the dynamics of a single charged 
particle moving at a constant velocity
above the surface of a dielectric fluid. 
Here the cooperative behaviour emerges 
because of the
electrostatic interactions between 
the particle and the fluid molecules.  
The
particle perturbs the fluid surface creating  
a "bump", which travels together with the particle increasing 
effectively its mass
and thus producing a frictional 
drag force exerted on the particle.

\vspace{0.5cm}

The authors thank 
Profs. A.M.Cazabat, S.F.Burlatsky and E.Rapha\"el 
for helpful discussions.  Financial
 support from  the COST Project  D5/0003/95
  and the FNRS
 is gratefully
acknowledged.

\vspace{0.8cm}

Fig.1. A schematic picture of a liquid 
monolayer on solid surface. 
The black circle denotes the probe
molecule and smaller grey circles represent 
the monolayer particles.
 Wavy lines depict the wafer-like
potential landscape created by the
 particle-solid interactions.

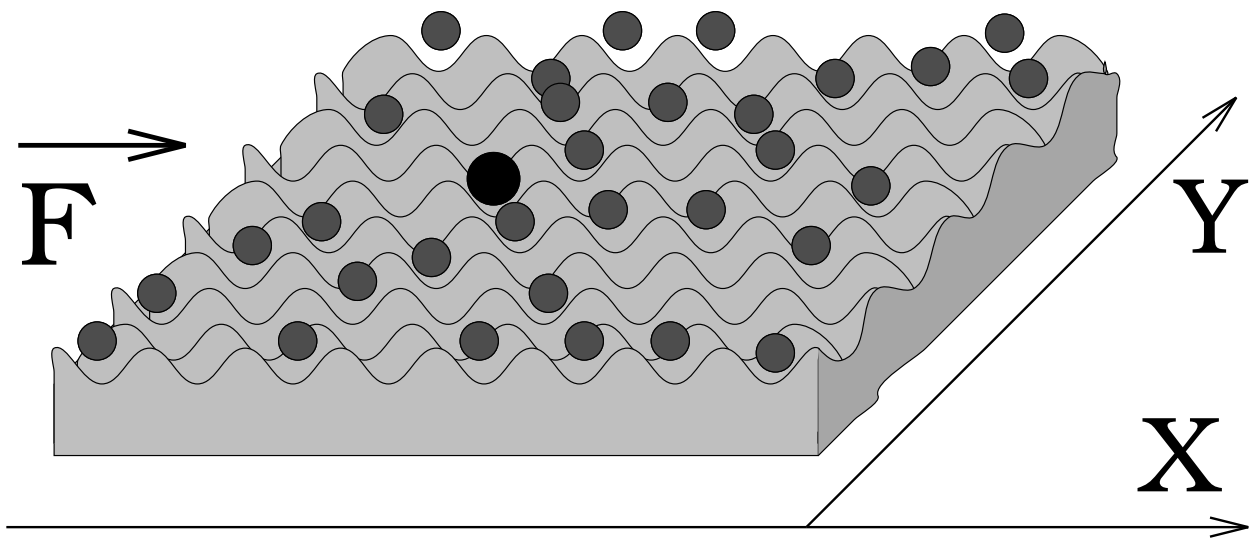
\end{document}